\documentstyle[prd,aps,epsfig]{revtex}
\newcommand{\be}{\begin{equation}}
\newcommand{\ee}{\end{equation}}
\newcommand{\ba}{\begin{eqnarray}}
\newcommand{\ea}{\end{eqnarray}}

\def\a{\alpha}
\def\b{\beta}
\def\d{\delta}
\def\e{\epsilon}

\def\f{\phi}
\def\vf{\varphi}
\def\g{\gamma}
\def\h{\eta}
\def\j{\psi}

\def\l{\lambda}
\def\m{\mu}
\def\n{\nu}
\def\p{\pi}
\def\q{\theta}

\def\x{\xi}
\def\z{\zeta}

\def\S{\Sigma}

\def\cm{{\cal M}}

\def\cv{{\cal V}}

\newcommand{\ov}{\overline}

\newcommand{\dg}{\dagger}
\newcommand{\pr}{\prime}
\newcommand{\aand}{\;\;\;\mbox{and}\;\;\;}
\newcommand{\pa}{\partial}
\def\sl#1{\rlap{\hbox{$\mskip 1 mu /$}}#1}

\newcommand{\ra}{\rangle}
\newcommand{\la}{\langle}
\newcommand{\sx}{\sigma_x}
\newcommand{\sy}{\sigma_y}
\newcommand{\sz}{\sigma_z}

\begin{document}

\draft
\title{Electron-electron attractive interaction in Maxwell-Chern-Simons QED$_3$ 
at zero temperature}
\author{
H. Belich$^{a}$\thanks{\tt belich@cbpf.br}, 
O.M. Del Cima$^{b}$\thanks{\tt delcima@gft.ucp.br}, 
M.M. Ferreira Jr.$^{a,c}$\thanks{\tt manojr@cbpf.br} and 
J.A. Helay\"el-Neto$^{a,b}$\thanks{\tt helayel@gft.ucp.br}     }
\address{$^a${\it Centro Brasileiro de Pesquisas F\'\i sicas (CBPF),\\
Coordena\c c\~ao de Teoria de Campos e Part\'\i culas (CCP), \\
Rua Dr. Xavier Sigaud 150 - 22290-180 - Rio de Janeiro - RJ - Brazil.}\\
$^b${\it Grupo de F\'\i sica Te\'orica (GFT), \\ 
Universidade Cat\'olica de Petr\'opolis (UCP), \\ 
Rua Bar\~ao do Amazonas 124 - 25685-070 - Petr\'opolis - RJ - Brazil.}\\
$^c${\it Universidade Federal do Maranh\~ao (UFMA), \\ 
Departamento de F\'\i sica, \\ 
Campus Universit\'ario do Bacanga - 65085-580 - S\~ao Luiz - MA - Brazil.}   }

\date{\today}
\maketitle

\begin{abstract}
One discusses the issue of low-energy electron-electron bound states in the 
Maxwell-Chern-Simons model coupled to QED$_3$ with spontaneous
breaking of a local $U(1)$-symmetry. The scattering potential, in the non-relativistic 
limit, steaming from the electron-electron M{\o}ller scattering, mediated by the 
Maxwell-Chern-Simons-Proca gauge field and the Higgs scalar, might be attractive by 
fine-tuning properly the physical parameters of the model.  
\end{abstract}
\pacs{PACS numbers: 11.10.Kk 11.15.-q 11.15.Ex 11.80.-m 
\hspace{4,05cm}ICEN-PS-01/01}
\centerline{\tt In memory of George Leibbrandt}
\section{Introduction}
In the latest years, planar QED (QED$_{3}$) 
\cite{djt,kogan,girotti,hagen,dobroliubov,szabo,groshev,georgelin,delcima,phdthesis,kogan1
} 
has been
object of intensive investigation, both by its formal aspects and by the
possibilities of application to important phenomena in the realm of 
Condensed Matter Physics, namely high-$T_{\rm c}$ superconductivity and quantum
Hall effect. The first phenomenon, as well-known, is related to the existence of
electron-electron bound states, but the nature of these pairs belonging to
high-$T_{\rm c}$ phase is not still set up. The absence of a ultimate theory for 
high-$T_{\rm c}$ superconductivity has attracted attention
of a large number of condensed matter and field theorists.

The search for a mechanism inducing the formation of electron-electron bound states has 
also
passed through QED$_{3}$, since high-$T_{\rm c}$ superconductivity is supposed
to be a quasi-planar phenomenon. Moreover, it is known that the Coulombian
interaction in three space-time dimensions leads to a confining potential rather than a
condensating one, which indicates the necessity of a finite range
interaction. One should stress here that, in spite of some claims found out in the 
literature, the electromagnetic potential cannot be of the $1/r$-type in three space-time 
dimensions, for it would demand a highly non-local action, leading to serious troubles as 
long as causal propagation of particles is concerned; on the other hand, it does not lead 
to bound states \cite{chadan}, contrary to what happens in four space-time dimensions. 
The idea of providing mass to the photon was then proposed as an
attempt to try to by-pass this difficulty. In this sense, the
Maxwell-Chern-Simons (MCS) model \cite{djt} was adopted as a mechanism for generating 
(topological) mass for the photon. A deconfining potential consequently emerges and the 
quest 
for electron-electron bound states turns out to be a sensible matter. All these aspects 
have been 
embrassed in \cite{kogan}, where the MCS model coupled to QED$_3$ is considered as a main 
tool for investigation of fermion-fermion scattering processes mediated by a topological 
massive gauge boson. The issue of electron-electron bound states, in the MCS QED$_3$, has 
been taken into account for the first time by numerical simulations in  
Ref.\cite{girotti}, however, 
their result is characterized by the fact that just one-photon exchange diagrams have 
been taken into account, leading to an incomplete Aharonov-Bohm potential term 
\cite{hagen}. 
The authors of Ref.\cite{dobroliubov} comment on the results presented in \cite{girotti} 
asserting that they hold on for small $k$ (statistics parameter), nevertheless in this 
regime perturbation theory breaks down and higher order
contributions to the electron-electron scattering amplitude become also important, so that 
the term in $1/k^2$, stemming from the two-photon exchange diagrams, could not be 
neglected. 
The solution to this controversy consists in
considering the two-photon exchange diagrams \cite{dobroliubov,szabo}, whose contribution
to the order $1/k^2$ to the scattering potential restores the gauge invariance in the 
non-relativistic limit \cite{hagen1} of the
theory and circumvents the erroneous conclusion of an attractive centrifugal
barrier. Indeed, the work of Ref.\cite{dobroliubov} displays a number of interesting 
limits where $e^-e^-$ and $e^+e^+$ bound state formation may be analyzed and the important 
outcome is that the Pauli dipole interaction among the electrons, due to their magnetic 
moment, may, in a suitable limit, dominate over the charge-charge repulsive interaction, 
leading to bound state formation, which has been also addressed in \cite{groshev}. It is 
also concluded that, in the case of light gauge bosons, the MCS model minimally coupled to 
QED$_3$ does not provide electron-electron bound states. The MCS model non-minimally 
coupled to fermions and bosons carrying an anomalous magnetic moment and within the 
perturbative
region $1/k\ll1$ has been analyzed in \cite{georgelin}. The presence of this non-minimal 
coupling is pointed out to be a key factor for the appearance of an attractive potential 
between charges of same sign.

Until the present moment, one observes that all the quoted works concerning 
electron-electron bound states make use of the Chern-Simons term as the only
mechanism yielding the photon mass. In our work, one employs a different
theoretical approach to generate photon mass (beyond the topological one)
and possibly the electron-electron bound states. Specifically, one adopts a
Maxwell-Chern-Simons model minimally coupled to QED$_{3}$ with spontaneous breaking of a 
local
$U(1)$-symmetry. Similarly, in Refs.\cite{delcima,phdthesis}, the Higgs mechanism has been 
used in the framework of a parity-preserving QED$_3$ in searching for electron-electron 
bound states. The symmetry breaking is accomplished by a
sixth-power potential, where a Higgs scalar and a
massive gauge boson (Maxwell-Chern-Simons-Proca) stem as a by-product from the breaking of 
a local $U(1)$-symmetry. As we shall present here, the low-energy M{\o}ller scattering
mediated by these two quanta points to the real possibility of an
attractive $e^-$-- $e^-$ scattering potential.
Thus, it becomes manifest that the Higgs mechanism has the relevant role of
allowing electron-electron pair condensation. In fact, our proposal is based upon the 
Higgs exchange to bind the electron pair rather than on a mass relationship that leads to 
a dominance of magnetic moment interaction over the charge repulsion.

Our paper is organized as follows. In Section II, the Maxwell-Chern-Simons model coupled 
to QED$_3$ with spontaneous breaking of a $U(1)$-symmetry is introduced. The low-energy 
electron-electron scattering potential, in the Born approximation, is derived and 
discussed in Section III. The final conclusions are left to Section IV. In the Appendix, 
Section V, general physical properties of planar QED are alucidated.  

\section{The Maxwell-Chern-Simons QED and the Higgs mechanism}

The action for the Maxwell-Chern-Simons model coupled to QED$_3$ with a local 
$U(1)$-symmetry  is given by:
\ba
S_{\rm QED}=\int d^3x\biggl\{\!\!\!&-&\frac14 F^{\m\n}F_{\m\n}+i\ov\j \g^\m D_\m\j+
\frac12 \q\e^{\m\n\a}A_\m\pa_\n A_\a-m_e\ov\j \j-y\ov\j \j\vf^*\vf+D^\m\vf^*D_\m\vf + 
\nonumber\\
&-& V(\vf^*\vf)\biggr\}~, \label{action1}
\ea
where the $V(\vf^*\vf)$ is a sixth-power potential, being the most general renormalizable 
$U(1)$-invariant potential in three dimensions \cite{delcima,phdthesis}:
\be
V(\vf^*\vf)=\m^2\vf^*\vf+\frac\z2 (\vf^*\vf)^2+\frac\l3 (\vf^*\vf)^3~.
\ee
The covariant derivatives are defined as follows:
\be
D_\m\j=(\pa_\m+ieA_\m)\j \aand D_\m\vf=(\pa_\m+ieA_\m)\vf~.
\ee
In the action S$_{\rm QED}$, Eq.(\ref{action1}), $F_{\m\n}$ is the usual field strength 
for $A_\m$, $\j$
is a spinor field describing a fermion with positive spin polarization (spin up) and an 
anti-fermion with negative spin polarization (spin down) \cite{delcima,phdthesis},
whereas $\vf$ is a complex scalar field. In three space-time dimensions, the positive- and 
negative-energy solutions have their polarization fixed by the signal of mass in the Dirac 
mass term \cite{delcima,phdthesis,binegar}. The conventions\footnote{The metric adopted is 
$\h^{\m\n}={\rm diag(+,-,-)}$ and the $\g$-matrices are taken as 
$\g^\m=(\sz,i\sx,-i\sy)$.} adopted here are
stated in the Appendix, where the mass dimensions of all the fields and parameters are 
displayed in the Table~\ref{table2}.

The sixth-power potential is the responsible for breaking the electromagnetic
$U(1)$-symmetry. Analyzing the structure of the potential $V(\vf^*\vf)$, one must impose 
that it is bounded from below and it yields only
stable vacua (metastability is ruled out). These requirements reflect on the
following conditions on the parameters $\m$, $\z$ and $\l$ \cite{delcima,phdthesis}:
\be
\l>0~,~~\z<0 \aand \m^2\leq\frac{3\z^2}{16\l}~.
\ee

Considering $\langle\vf^*\vf\rangle=v^2$, the vacuum expectation value for the
scalar field product $\vf^*\vf$ is given by
\be
\langle\vf^*\vf\rangle=v^2=-\frac{\z}{2\l} + \sqrt{\left(\frac{\z}{2\l}\right)^2 - 
\frac{\m^2}{\l}}~,
\ee
while the minimum condition reads 
\be
\m^2+\z v^2+\l v^4=0~.
\ee

In order to preserve the manifest renormalizability of the model, one adopts the 't Hooft 
gauge:
\be
S_{{\rm R}_\x}=\int d^3x \biggl\{ -\frac{1}{2\x}(\pa^\m A_\m-\sqrt{2}\x 
M_A\chi)^2\biggr\}~.
\ee
Then, by adding it up to the action (\ref{action1}), and assuming the following 
parametrization for the scalar field,  
\be
\vf=v+H+i\chi~,
\ee 
where $H$ represents the Higgs scalar and $\chi$ the would-be Goldstone boson, the 
Maxwell-Chern-Simons QED$_3$ action with the $U(1)$-symmetry spontaneously broken is as 
follows 
\ba
S_{\rm QED}^{\rm broken}=\int d^3x \biggl\{\!\!\!&-&\frac14 F^{\m\n}F_{\m\n}+\frac12 M_A^2 
A^\m A_\m - \frac1{2\x}(\pa^\m A_\m)^2 + \ov\j(i\g^\m D_\m - m)\j + \frac12 \q\e^{\m\n\a} 
A_\m\pa_\n A_\a + \nonumber\\
&+&\pa^\m H \pa_\m H + {\pa^\m}\chi {\pa_\m}\chi - \x M^2_A\chi^2 - y \ov\j \j (2vH + 
H^2+\chi^2) + 2eA^\m(H{\pa_\m}\chi - \chi{\pa_\m}H) + 
\nonumber\\
&+& e^2 A^\m A_\m(2vH+H^2+\chi^2) - \m^2((v+H)^2+\chi^2) - {\z\over2}((v+H)^2+\chi^2)^2 - 
{\l\over3}((v+H)^2+\chi^2)^3 \biggr\}~,\label{action2}
\ea 
where the mass parameters $M^2_A$, $m$ and $M^2_H$, read
\be
M^2_A=2v^2e^2~,~~m=m_e+yv^2 \aand M^2_H=2v^2(\z + 2 \l v^2)~.\label{masses}
\ee

\section{The low-energy electron-electron scattering potential}
The issue of electron-electron bound states in the Maxwell-Chern-Simons model coupled to 
planar QED has been addressed to in the literature since the end of the eighties 
\cite{kogan,girotti,hagen,dobroliubov}, motivated by possible applications to the 
parity-breaking high-$T_{\rm c}$ superconductivity phenomenon.

In this Section, we shall present the evaluation of the electron-electron scattering 
potential in the low-energy approximation. The M{\o}ller electron-electron scattering 
process is mediated by the Higgs scalar and the Maxwell-Chern-Simons-Proca gauge field. 
In order to compute the scattering potential through the M{\o}ller electron-electron 
amplitude, we show the propagators associated to the Higgs ($H$), the fermion ($\j$) and 
the massive gauge boson ($A_\m$), which stem straightforwardly from the action 
(\ref{action2}), as presented below
\ba
&&\langle\ov\j(k)\j(k)\rangle=i\frac{{\sl k}+m}{k^2-m^2}~,~~
\langle H(k)H(-k)\rangle=\frac{i}2 \frac{1}{k^2-M_H^2} \aand \nonumber\\
&&\langle A_\m(k)A_\n(-k)\rangle=-i\biggl\{\frac{k^2-M_A^2}{(k^2-M_A^2)^2-k^2
\q^2}\biggl(\h_{\m\n}-\frac{k_\m k_\n}{k^2}\biggr)+\frac{\x}{(k^2-\x M_A^2)}\frac{k_\m 
k_\n}{k^2}+\frac{\q}{(k^2-M_A^2)^2-k^2\q^2}i\e^{\m\a\n}k_\a\biggr\}~.
\ea

The propagator of the Maxwell-Chern-Simons-Proca field given above can be rewritten 
in the following way
\ba
\langle A_\m(k)A_\n(-k)\rangle =&-&i\biggl\{ 
\biggl[\frac{C_+}{k^2-M_+^2}+\frac{C_-}{k^2-M_-^2}\biggr]\biggl(\h_{\m\n}-\frac{k_\m 
k_\n}{k^2}\biggr)+\frac{\x}{(k^2-\x M_A^2)}\frac{k_\m k_\n}{k^2} + \nonumber\\
&+&\left[\frac{C}{k^2-M_+^2}-\frac{C}{k^2-M_-^2}\right]i\e_{\m\a\n}k^\a \biggr\}~,
\ea
where the positive definite constants $C_{+}$, $C_{-}$, $C$, and the squared masses 
$M^2_+$ 
and $M^2_-$, are given by:
\be
C_\pm=\frac12\biggl[1 \pm \frac{\q}{\sqrt{4M_A^2+\q^2}}\biggr]~,~~
C=\frac{1}{\sqrt{4M_A^2+\q^2}}~,
\ee
\be
M_\pm^2=\frac{1}{2}\left[2M_A^2+\q^2\pm |\q|\sqrt{4M_A^2+\q^2}\right]~,
\ee
with the massive poles, $M^2_+$ and $M^2_-$, corresponding to the two massive propagating 
quanta. It can be readily checked that both of them are physical states in that 
the residues at the poles are positive-definite.
From the action $S_{\rm QED}^{\rm broken}$, given by Eq.(\ref{action2}), it can be derived 
the Feynman rules associated to the electromagnetic and Yukawa interactions, 
$\cv_{\j H \j}=2ivy$ and $\cv_{\j A \j}=ie\g^\m$, respectively. 

\begin{figure}[h]
\vspace{-9,1cm}
\epsfxsize=15cm
\centerline{\epsfbox{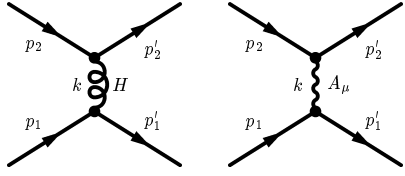}} 
\vspace{-15cm}
\caption{Feynman diagrams associated to the $e^-$-- $e^-$ scattering mediated by the Higgs 
and the massive gauge boson.}
\label{graphs}
\end{figure}

Let us now start the derivation of the electron-electron scattering potential through 
the total M{\o}ller scattering amplitude ($\cm_{\rm total}$) in the low-energy 
approximation, {\it i.e.}, the non-relativistic limit ($\cm_{\rm total}^{\rm nr}$). 
The scattering potential is nothing but the two-dimensional Fourier transform of the 
lowest-order $\cm_{\rm total}$-matrix element:
\be
V(r)=\int \frac{d^2{\vec k}}{(2\p)^2}~\cm_{\rm total}^{\rm nr}~e^{i{\vec k}\cdot
{\vec r}}~.\label{fourier}
\ee
In the case we are analyzing here (the electron-electron scattering, $e^-$-- $e^-$, 
~mediated 
by the Higgs, $H$, ~and the massive gauge boson, $A_\m$), the matrix 
$\cm_{\rm total}^{\rm nr}$ that appears in Eq.(\ref{fourier}) is precisely the 
part of the covariant matrix element which corresponds to direct scattering, $s$-channel. 
This can be understood in view of the fact that antisymmetric wave functions in 
non-relativistic quantum mechanics automatically take care of the contributions resulting 
from the exchange scattering. The $s$-channel amplitudes for the $e^-$-- $e^-$ scattering 
mediated by the Higgs and the gauge field, with the corresponding Feynman diagrams 
displayed in Fig.\ref{graphs}, are listed below:
\begin{enumerate}
\item Scattering amplitude with the Higgs exchange:
\be
-i\cm_{e^-He^-}=\ov{u}(p_1)(2ivy)u(p^\pr_1)\langle H(k)H(-k)\rangle 
\ov{u}(p_2)(2ivy)u(p^\pr_2)~,\label{ehe}
\ee
\item Scattering amplitude with the massive gauge boson exchange:
\be
-i\cm_{e^-Ae^-}=\ov{u}(p_1)(ie\g^\m)u(p^\pr_1)\langle A_\m(k)A_\n(-k)\rangle  
\ov{u}(p_2)(ie\g^\n)u(p^\pr_2)~,\label{eae}
\ee
\end{enumerate}
where $k^2=(p^\pr_1-p_1)^2$ is the invariant squared momentum transfer. In the partial 
scattering amplitudes $\cm_{e^-He^-}$ and $\cm_{e^-Ae^-}$, given by Eqs.(\ref{ehe}) 
and (\ref{eae}), respectively, the spinor $u(p)$ is the positive-energy solution of the 
Dirac equation for
$\j$, satisfying the following normalization condition stated in the Appendix: 
\be
\ov{u}(p)u(p)=1~.
\ee
The momenta configuration in the center-of-mass frame (c.m.) of the two interacting 
electrons, 
as well as the momentum transfer, are chosen as
\ba
&p_1=(E,p,0)~,~~p^\pr_1=(E,p\cos\f,p\sin\f)~,\nonumber\\
&p_2=(E,-p,0)~,~~p^\pr_2=(E,-p\cos\f,-p\sin\f) \aand \nonumber\\
&k=p^\pr_1-p_1=(0,p(\cos\f-1),p\sin\f)~,\label{pconfig}
\ea
where $\f$ is the c.m. scattering angle, which is defined as the angle between the 
directions 
in the center-of-mass frame of the two incoming (initial state) and outgoing (final state) 
electrons.
Assuming the momenta configuration above (\ref{pconfig}), the total scattering amplitude 
in the low-energy approximation, $\cm_{\rm total}^{\rm nr}$, can now be derived from 
the partial ones, $\cm_{e^-He^-}^{\rm nr}$ and $\cm_{e^-Ae^-}^{\rm nr}$: 
\be
\cm_{\rm total}^{\rm nr}=\cm_{e^-He^-}^{\rm nr} + \cm_{e^-Ae^-}^{\rm nr}~,
\ee
where
\be
\cm_{e^-He^-}^{\rm nr}=-2v^2y^2\frac{1}{\vec{k}^2 + M_H^2}~,
\ee
and
\ba
\cm_{e^-Ae^-}^{\rm nr}&=&\cm_1 + \cm_2 + \cm_{\rm AB} \nonumber\\
&=& e^2\left[\frac{C_+}{\vec{k}^2 + M_+^2} + \frac{C_-}{\vec{k}^2 + M_-^2}\right] + 
e^24\frac{p^2}{2m}(1-\cos\f)\left[\frac{C}{\vec{k}^2 + M_+^2} - 
\frac{C}{\vec{k}^2 + M_-^2}\right] + \nonumber\\ 
&+& ie^24\frac{p^2}{2m}\sin\f\left[\frac{C}{\vec{k}^2 + M_+^2} - 
\frac{C}{\vec{k}^2 + M_-^2}\right]~.\label{ampa}
\ea
Notice that the first two terms of the massive gauge field amplitude, 
$\cm_{e^-Ae^-}^{\rm nr}$, given in Eq.(\ref{ampa}), $\cm_1$ and $\cm_2$, are the 
real part of the M{\o}ller scattering amplitude, whereas the last one, $\cm_{\rm AB}$, 
which is imaginary, is the Aharonov-Bohm amplitude for the fermions 
\cite{kogan,dobroliubov,georgelin,kogan1}. The total M{\o}ller scattering amplitude 
in the non-relativistic limit reads as below:
\ba
\cm_{\rm total}^{\rm nr}=&-&2v^2y^2\frac{1}{\vec{k}^2 + M_H^2} + 
e^2\left[\frac{C_+}{\vec{k}^2 + M_+^2} + \frac{C_-}{\vec{k}^2 + M_-^2}\right] + 
e^24\frac{p^2}{2m}(1-\cos\f)\left[\frac{C}{\vec{k}^2 + M_+^2} - 
\frac{C}{\vec{k}^2 + M_-^2}\right] + \nonumber\\ 
&+& ie^24\frac{p^2}{2m}\sin\f\left[\frac{C}{\vec{k}^2 + M_+^2} - 
\frac{C}{\vec{k}^2 + M_-^2}\right]~.\label{ampt}
\ea

Now, bearing in mind that the non-relativistic scattering potential in the Born 
approximation is obtained from the scattering amplitude (\ref{ampt}) through the 
Fourier transform given by Eq.(\ref{fourier}), one gets:
\ba
V(r)&=&V_{\rm Higgs}(r) + V_{\rm gauge}(r)~,\\
V_{\rm Higgs}(r)&=&-\frac{1}{2\p}2v^2y^2K_0(M_Hr)~,\\
V_{\rm gauge}(r)&=&V_1(r) + V_2(r) + V_{\rm AB}(r) \nonumber\\
&=&\frac{e^2}{2\p} \left[C_+K_0(M_+r) + C_-K_0(M_-r)\right] - \frac{e^2}{2\p} \frac{C}{m} 
\left[M_+^2K_0(M_+r)-M_-^2K_0(M_-r)\right] + \nonumber\\ 
&+&2\frac{e^2}{2\p}\frac{l}{mr}C\left[M_+K_1(M_+r) - M_-K_1(M_-r)\right]~.
\ea
Therefore, the $e^-$-- $e^-$ low-energy scattering potential, $V(r)$, is given by
\ba
V(r)=&-&{\frac{1}{2\p}} 2v^2y^2K_0(M_Hr) + \frac{e^2}{2\p}\biggl\{\left[C_+ - 
\frac{C}{m}M_+^2\right]K_0(M_+r) + \left[C_- + \frac{C}{m}M_-^2\right]K_0(M_-r) + 
\nonumber\\
&+&2\frac{l}{mr}C\left[M_+K_1(M_+r) - M_-K_1(M_-r)\right]\biggr\}~, \label{V}
\ea
where, $K_0$ and $K_1$ are the zeroth- and first-order modified Bessel functions of the 
second kind, respectively, and $l$ is the angular momentum. 

It should be stressed here that the low-energy electron-electron scattering potential
we are deriving is valid only in the perturbative regime, where loop corrections are 
negligible if compared to the semi-classical approximation. Perturbation theory is 
realized 
whenever dimensionless parameters are kept much smaller than one. At the broken-symmetry 
phase, the Maxwell-Chern-Simons model coupled to planar QED has four dimensionless 
parameters, 
${e^2}/m$, ${e^2}/{M_H}$, ${e^2}/{M_+}$ and ${e^2}/{M_-}$. Nevertheless, the masses $M_H$ 
and $M_-$ vanish in the unbroken-symmetry phase (when $v^2\rightarrow 0$), in this way 
${e^2}/m$ and ${e^2}/{M_+}$ remain the natural dimensionless parameters respect to which 
perturbation theory shall be performed. For our purposes here (where the low-energy 
electron-electron potential is derived through the Born approximation of the M{\o}ller 
scattering amplitude in the non-relativistic limit), since the electron is the heaviest 
particle (electron effective mass (\ref{masses}), $m\approx 0,5 MeV$) with the Higgs 
(in condensed matter phenomena, $M_H\approx meV$) and the massive gauge boson 
($M_\pm\approx meV$) being the intermediate quanta, we can ensure confidence on 
the perturbative regime by assuming ${e^2}/m$ and $y\ll1$ provided that ${e^2}/{M_+}\ll1$.

Non-trivial aspects of the Galilean (non-relativistic) limit of a gauge theory are 
discussed 
in the work of Hagen \cite{hagen1}. In the non-relativistic limit, even though the 
perturbative regime is considered, besides the one-photon exchange diagrams, one has to 
take into account two-photon exchange contributions so as to preserve gauge invariance 
(the non-relativistic Hamiltonian is quadratic in momentum), as presented by the authors 
of Refs.\cite{dobroliubov,szabo} in the framework of a Maxwell-Chern-Simons model 
minimally coupled either to fermions or to fermions and scalars. The non-relativistic 
scattering potential for the MCS QED$_3$ model has been derived in Ref.\cite{dobroliubov}, 
there the perturbative regime is established by the statistics parameter $k$ 
(in our case it is given by $4\p\q/e^2$) whenever $1/k\ll1$. In order to guarantee 
gauge invariance in the low-energy approximation, despite of $1/k\ll1$, two-photon 
exchange diagrams have to be taken into account as well, which leads to the correct 
low-energy electron-electron scattering potential for the MCS model coupled to planar 
QED \cite{dobroliubov} as follows:
\be
V_{\rm MCS}(r)=\frac{e^2}{2\p}\left[1-\frac{\q}{m}\right]K_{0}(\q r)+\frac{1}{mr^2}
\left\{l-\frac{e^2}{2\p\q}[1-\q rK_1(\q r)]\right\}^2 ~.\label{Vmcs}
\ee
For feasible applications to Condensed Matter Physics, which should require $\q\ll m$, the 
non-relativistic MCS QED$_3$ scattering potential, given above by Eq.(\ref{Vmcs}), results 
to be repulsive, where its first term corresponds to the electromagnetic potential whereas 
the last one includes the Aharonov-Bohm, the centrifugal barrier and the two-photon 
exchange 
contributions. 

Let us now remind that our main task is to derive the gauge-invariant scattering potential 
in the non-relativistic limit for the model proposed here. In this way, this amounts in 
adding to Eq.(\ref{V}) the centrifugal barrier and the one-loop corrections resulting from 
the two-photon exchange diagrams, by following the steps pointed out in 
Refs.\cite{dobroliubov,szabo} together with the general arguments on non-relativistic 
gauge theories analyzed in \cite{hagen1}. Therefore, as a final result, the 
non-relativistic 
effective scattering potential of the MCS QED$_3$ model with spontaneous symmetry 
breaking, 
$V_{\rm eff}(r)$, reads as below: 
\ba
V_{\rm eff}(r)=&-&{\frac{1}{2\p}} 2v^2y^2K_0(M_Hr) + \frac{e^2}{2\p}\left\{\left[C_+ - 
\frac{C}{m}M_+^2\right]K_0(M_{+}r) +
\left[C_- + \frac{C}{m}M_-^2\right]K_0(M_-r)\right\} + \nonumber\\
&+&\frac{1}{mr^2}\left\{l+\frac{e^2}{2\p}Cr[M_+K_1(M_+r) - M_-K_1(M_-r)]\right\}^2 ~, 
\label{Veff}
\ea
where $\frac{l^2}{mr^2}$ is the centrifugal barrier and
the term in $C^{2}$ arises from the one-loop two-photon exchange diagrams 
\cite{dobroliubov,szabo}. It can be concluded from the effective electron-electron 
scattering potential $V_{\rm eff}(r)$, that the only attractive contribution to it comes 
from the Higgs interaction given by the first term in Eq.(\ref{Veff}). However, the second 
term, which is proportional to $e^2/2\p$, shows to be repulsive in the range of parameters 
we are restricting our model, whereas the last one has always the same behavior, namely, 
repulsive. In view of the attractive nature of the Yukawa interaction, by an appropriate 
fine-tuning of the parameters (coupling constants and masses) of the model, so as to 
compensate the repulsion caused by the electromagnetic interaction and the ``effective'' 
centrifugal barrier, the M\o ller scattering potential $V_{\rm eff}(r)$ turns out to be 
attractive. As a consequence, this might favor electron-electron bound states provided 
$V_{\rm eff}(r)$ is ``weak'' in the sense of Kato and satisfies the Set\^o bound 
as discussed by Chadan {\it et al.} \cite{chadan} in the framework of low-energy 
scattering 
in three space-time dimensions, this issue is now under investigation \cite{boundmcsqed}. 

\section{General Conclusions}
The low-energy electron-electron scattering potential we have derived for the MCS QED$_3$ 
model with spontaneous symmetry breaking sets up the physical framework for the
mechanism of an electron-electron pairing and the consequent formation of bound
states. The Higgs contribution to the effective scattering potential reveals to be always 
attractive while the gauge boson contribution is repulsive in the range of parameters 
dictated by the condensed matter phenomena, namely, $\q\ll m$. Therefore, one concludes 
that the $e^-$-- $e^-$ scattering potential, $V_{\rm eff}(r)$, given by Eq.(\ref{Veff}), 
is always attractive whenever, by a properly fine-tuning of the parameters, the attraction 
caused by the Higgs mediation becomes stronger than the repulsion yields 
by the gauge field mediation and the ``effective'' centrifugal barrier.
Thus, as a conclusion, the Higgs mechanism \cite{delcima,phdthesis} provides a
possible mechanism for an electron-electron attractive potential, and therefore
sets up an effective possibility for pair condensation at the low-energy
limit of a parity-breaking QED$_{3}$. Finally, one points out that this model
bypasses the difficulties found by several authors \cite{girotti} who tried
to obtain electron-electron bound states in MCS QED$_{3}$ by only considering the exchange 
of gauge bosons.

It is important to observe that the gauge-mediated contribution, $V^{\rm gauge}_{\rm 
eff}(r)$ (the last three terms of Eq.(\ref{Veff})), to the scattering potential, $V_{\rm 
eff}(r)$, 
\ba
V^{\rm gauge}_{\rm eff}(r)&=&\frac{e^2}{2\p}\left\{\left[C_+ - 
\frac{C}{m}M_+^2\right]K_0(M_{+}r) +
\left[C_- + \frac{C}{m}M_-^2\right]K_0(M_-r)\right\} + \nonumber\\
&+&\frac{1}{mr^2}\left\{l+\frac{e^2}{2\p}Cr[M_+K_1(M_+r) - M_-K_1(M_-r)]\right\}^2 ~, 
\label{Veffgauge}
\ea
reproduces the usual form for a vanishing Proca photon mass. In this limit
\be
M_+\longrightarrow\q~,~~M_-\longrightarrow 0~,~~C_+\longrightarrow 1~,~~
C_-\longrightarrow 0~,~~K_1(M_-r)\longrightarrow\frac{1}{M_-r}~,~~
C\longrightarrow\frac{1}{\q}~,
\ee
such that one has
\be
\lim_{M_A\longrightarrow 0}V^{\rm gauge}_{\rm 
eff}(r)=\frac{e^2}{2\p}\left[1-\frac{\q}{m}\right]
K_{0}(\q r) + \frac{1}{mr^2}\left\{l-\frac{e^2}{2\p\q}[1-\q rK_{1}(\q r)]\right\}^2~,
\ee
which is exactly the same as the one obtained in the works of 
Refs.\cite{kogan,dobroliubov}.

To conclude, we would like to stress that we shall next check whether or not low-energy 
electron-electron bound states stem from the MCS QED$_3$ model with spontaneous symmetry 
breaking. This shall be done by explicitly solving the Schr\"odinger equation with the 
help of 
numerical methods. Our results shall be reported elsewhere in a forthcoming paper 
\cite{boundmcsqed}.

\section*{Acknowledgements} 
The authors would like to thank Olivier Piguet for helpful discussions. 
One of the authors (O.M.D.C.) dedicates this work to 
his wife, Zilda Cristina, to his daughter, Vittoria, to his son 
Enzo, and to his mother, Victoria. He also dedicates it to the memory of George Leibbrandt 
for the example of physicist has shown to be. Paz Profunda George!

\section{Appendix}

Here we present some aspects of a massive Dirac spinor 
living in three space-time dimensions, like the positive and negative energy
solutions 
to the Dirac equation satisfied by $\j$. We present the 
Hamiltonian for $\j$, and also compute 
explicitly the
charges of the positive and negative energy wave functions 
associated to $\j$.

\subsection{Positive and negative energy solutions 
for $\j$}
Let us consider $u$ and $v$, respectively, as the positive 
and negative solutions to the Dirac equations for 
$\j$. Therefore, they satisfy the following equations in momentum space:
\be
({\sl{p}} - m) u(p) = 0 \aand (-{\sl{p}} - m) v(p) = 0~. \label{Diraceq} 
\ee
Their solutions are given by
\be
u(p)={ {{\sl{p}} + m}\over{\sqrt{2m(m+E)}} }
~u(m,\vec{0})\aand
v(p)={ -{{\sl{p}} + m}\over{\sqrt{2m(m+E)}} }
~v(m,\vec{0})~, \label{gensolu}
\ee
where $E\equiv k^0={\sqrt{ {\vec{k}}^2 + m^2 }}>0$. 
The wave functions 
$u(m,\vec{0})$ and $v(m,\vec{0})$ are the solutions of 
Eqs.(\ref{Diraceq}) in the rest frame 
\be
u(m,\vec{0})=
\left(\begin{array}{c}
1\\
0
\end{array}\right) \aand
v(m,\vec{0})=
\left(\begin{array}{c}
0\\
1
\end{array}\right)~.\label{restsolu}
\ee

The positive and negative energy solutions given by 
Eqs.(\ref{gensolu}) are normalized to :
\be
{\ov u}(p) u(p)=1 \aand
{\ov v}(p) v(p)=-1~.\label{norm}
\ee

\subsection{The spin of $u$ and $v$}
Now, by considering the results of last subsection, 
one is able
to determine the spins of the solutions 
$u$ and $v$. We compute the spins in the 
particle rest frame, since we have in mind to explicitly 
exhibit the fact that the sign of the mass term fixes the 
polarization of the fermion.

In three space-time dimensions, the generators of the $\ov{SO(1,2)}$ group in 
the spinor representation read:
\be
\S^{kl}={\frac14}~[\g^k,\g^l]~,\label{skl}
\ee
where the $\g$-matrices are taken as
$\g^\m=(\sz,i\sx,-i\sy)$. 

The spin operator $S^{12}$ is obtained from (\ref{skl}), and 
it reads
\be
S^{12}={1\over 2}~\sz~.\label{spin}
\ee
Its action upon the rest frame wave functions 
given by Eqs.(\ref{restsolu}) is collected 
below:
\be
S^{12}u(m,\vec{0})=s^u u(m,\vec{0}) \aand
S^{12}v(m,\vec{0})=s^v v(m,\vec{0})~.
\label{eigens1} 
\ee

With the help of (\ref{restsolu}) and
(\ref{spin}), we find the following values for the spin 
eigenvalues 
$s^u$ and $s^v$: 
\be
s^u={1\over 2} \aand s^v=-{1\over 2}~.\label{spins}
\ee

An interesting point to stress here concerns the 
polarizations
of a particle ($u$) and the corresponding anti-particle 
($v$)
belonging to the same Dirac spinor ($\j$). As a typical
feature of 3 space-time dimensions, if a particle has 
spin $s$, its anti-particle has spin $-s$.  

\subsection{The Hamiltonian for $\j$}

Now, considering the Dirac equation for $\j$:
\be
(i{\sl{\pa} - m}) {\j} = 0~,
\ee
it follows that
\be
i{\pa\over{\pa t}}\j=\left(i\g^0\vec\g.
\vec{\pa} 
+ \b m\right)\j \equiv H_0\j~.
\ee
Therefore, for the general massive Dirac spinor, $\j$, the free 
Hamiltonian operator in momentum space, $H_0$, is given by:
\be
H_0 \j{\equiv}({\vec \a}.{\vec p} + \b m) \j~,\label{H0}
\ee
where  
\be
{\vec \a}=\g^0 {\vec \g} \aand \b=\g^0~.\label{ab}
\ee

\subsection{The spin of $u$ and $v$}

Let us consider the spin operator given by Eq.(\ref{spin}):
\be
S^{12}={1\over 2}~\sz~,\nonumber
\ee
and the free Hamiltonian 
operator in momentum space for the spinor $\j$ (Eq.(\ref{H0})): 
\be
H_0 \j{\equiv}({\vec \a}.{\vec p} + \b m) \j~,\nonumber
\ee
where ${\vec \a}$ and $\b$ are given by 
Eqs.(\ref{ab}). It can be easily shown that the 
following commutator vanishes
\be
\left[H_0,S^{12}\right]=0~.
\ee
This result ensures that the eigenvalues 
($s^u$ and $s^v$) of the spin operator, $S^{12}$, corresponding respectively to 
the wave functions $u$ and $v$ are indeed good quantum numbers 
to label physical states. 

\subsection{The charges of $u$ and $v$}

In order to determine the charges of the particles 
associated to the wave functions, $u$ and $v$, it is necessary to compute the 
eigenvalues of the charge operator, $Q$, respected to the field operator, $\j$. Its 
expansion in terms of the creation and annihilation operators reads as below:
\ba
&&\j(x) = \int{d^2\vec{k}\over (2\p)^2}{m\over k^0}
\left[a(k) u(k) e^{-ik.x} + b^{\dg}(k) v(k) 
e^{ik.x} \right]~,\label{exp1}\\
&&{\ov\j}(x) = \int{d^2\vec{k}\over (2\p)^2}{m\over k^0}
\left[a^{\dg}(k) {\ov u}(k) e^{ik.x} + 
b(k) {\ov v}(k) e^{-ik.x} \right]~,\label{exp2}
\ea
where the operators, $a^{\dg}$ and $b^{\dg}$, are the creation operators, and, $a$ and 
$b$, are the annihilation operators. 

With the help of the Dirac equation (\ref{Diraceq}), the normalization conditions 
(\ref{norm}) and the relation 
\be
\left\{ {\sl p},\g^0 \right\}=2p^0~,
\ee
the following equations are satisfied by the wave 
functions $u$ and $v$:
\ba
{u}^{\dg}(p) u(p)={p^0\over m} \aand
{v}^{\dg}(p) v(p)={p^0\over m}~.\label{norm1}
\ea

The microcausality fixes the following anticommutation 
relation:
\be
\left\{{\j}(x),{\j}^{\dg}(y) \right\}_{x^0=y^0}=
\d^2(\vec{x}-\vec{y})~.\label{micro}
\ee
Now, by assuming the field operator expansions 
(\ref{exp1}-\ref{exp2}), and the normalization 
condition given by Eq.(\ref{norm1}), 
the anticommutation relations between the creation and 
annihilation operators read:
\be
\left\{a(k),a^{\dg}(p) \right\}= (2\p)^2~{k^0\over m}~
\d^2(\vec{k}-\vec{p}) \aand
\left\{b(k),b^{\dg}(p) \right\}= (2\p)^2~{k^0\over m}
~\d^2(\vec{k}-\vec{p})~.\label{aa+bb+} 
\ee

The charge operator, $Q$, associated to 
the field operator, $\j$, is 
defined by the following normal ordering product:
\be
Q = \int{d^2\vec{x}}:j^0(x):=-e 
\int{d^2\vec{x}}:\j^{\dg}(x)\j(x):~,\label{j}
\ee
which in terms of the creation and annihilation 
operators are given by
\be
Q=-e \int{d^2\vec{k}\over (2\p)^2}{m\over k^0}
\left[a^{\dg}(k) a(k) -
b^{\dg}(k) b(k) \right]~. \label{Q}
\ee

From the anticommutation relations 
(\ref{aa+bb+}) and the Eq.(\ref{Q}), 
for the charge operator $Q$, it can be easily 
shown that
\be
\left[Q,a^{\dg}(p) \right]=-e~a^{\dg}(p) \aand
\left[Q,b^{\dg}(p) \right]=+e~b^{\dg}(p)~, \label{comm}
\ee

Let us denote the vacuum ground state by the ``ket'', 
$|0\ra$, such that
\be
a(k)|0\ra=0 \aand b(k)|0\ra=0~, 
\ee
where $\la0|0\ra=1$. Now, bearing in mind the commutation 
relations given by Eqs.(\ref{comm}), and 
applying them to the vacuum state, it follows that
\ba
&&Q|e^-\ra=-e~|e^-\ra 
~~~\mbox{where}~~~|e^-\ra = a^{\dg}|0\ra~;\\
&&Q|e^+\ra=+e~|e^+\ra 
~~~\mbox{where}~~~|e^+\ra = b^{\dg}|0\ra~. 
\ea
Due to these results, one concludes that:
\begin{enumerate}
\item{$a^{\dg}$ creates an electron ($u$) with spin 
$s^u={1\over 2}$ and charge $-e$.}
\item{$b^{\dg}$ creates a positron ($v$) with spin 
$s^v=-{1\over 2}$ and charge $+e$.}
\end{enumerate}

As a final conclusion, $u$ is a wave 
function of an electron ($e^-$) with spin $s^u={1\over 2}$, whereas $v$ is a wave function 
of a positron ($e^+$) with spin $s^v=-{1\over 2}$. Some of the physical 
relevant results obtained in this Appendix are summarized 
in Table~{\ref{table1}}.
\begin{table}
\begin{tabular}{|c|c|c|c|c|c|c|}
   Creation  &Charge  &Charge &Particle &Symbol 
&Wave &Spin  \\ 
   operator &operator & & & &function & \\
\hline
\hline
$a^{\dg}$ &$Q$ &$-e$ &electron &$e^-$ 
&$u$ &$s^u=+{1\over 2}$    \\
\hline
$b^{\dg}$ &$Q$ &$+e$ &positron &$e^+$ 
&$v$ &$s^v=-{1\over 2}$   \\
\end{tabular}
\caption{Charge and spin of the particles 
associated to the field operator ${\j}$.}\label{table1}
\end{table}
\begin{table}
\begin{tabular}{|c||c|c|c|c|c|c|c|c|c|c|}
& $A_\mu$ & $\j$ & $\vf$ & $m_e$ & $\q$ & $e$ & $y$ & $\m$ & $\z$ & $\l$ \\
\hline
\hline
$d$ & $1/2$ & $1$ & $1/2$ & $1$ & $1$ & $1/2$ & $0$ & $1$ & $1$ & $0$ \\ 
\end{tabular}
\caption{Mass dimensions of the fields and parameters.}\label{table2}
\end{table}

\end{document}